\documentclass[aps,twocolumn]{revtex4}
\usepackage{amsmath,amssymb}

\begin{document}

\title{Friedmann equations in braneworld scenarios from emergence of cosmic
space}
\author{A. Sheykhi $^{1,2}$\footnote{asheykhi@shirazu.ac.ir}, M. H. Dehghani $^{1,2}$ \footnote{mhd@shirazu.ac.ir} and S. E. Hosseini $^{1}$}
\address{$^1$  Physics Department and Biruni Observatory, College of
Sciences, Shiraz University, Shiraz 71454, Iran\\
         $^2$  Research Institute for Astronomy and Astrophysics of Maragha
         (RIAAM), P.O. Box 55134-441, Maragha, Iran}
\begin{abstract}
Recently, it was argued that the spacetime dynamics can be understood by
calculating the difference between the degrees of freedom on the boundary
and in the bulk in a region of space. In this Letter, we apply this new idea
to braneworld scenarios and extract the corresponding Friedmann equations of
$(n-1)$-dimensional brane embedded in the $(n+1)$-dimensional bulk with any
spacial curvature. We will also extend our study to the more general
Gauss-Bonnet braneworld with curvature correction terms on the brane and in
the bulk, and derive the dynamical equation in a nonflat Universe.
\end{abstract}

\maketitle

\section{Introduction\label{Intr}}

The emergence properties of gravity has a long history since the
original proposal made by Sakharov in 1968 \cite{Sak}. Recent
investigations supports the idea that gravitational field
equations in a wide range of theories can
be recast as the first law of thermodynamics on the boundary of space \cite%
{CaiKim,SheyW1,SheyW2,Shey0,Pad0}. Among various proposal on the connection
between thermodynamics and gravity, the so called entropic origin of gravity
proposed by Verlinde \cite{Ver}, has got a lot of attentions \cite%
{Cai4,Other,newref,sheyECFE,Ling,Modesto,Yi,Sheykhi2}. According to
Verlinde, gravity can be identified with an entropic force caused by changes
in the information associated with the positions of material bodies.
Verlinde considers the gravitational field equations as the equations of
emergent phenomenon and leaves the spacetime as a background geometric which
has already exist.

A new insight to the origin of spacetime dynamics, was recently suggested by
Padmanabhan\cite{Pad1} who claimed that the cosmic space is emergent as the
cosmic time progresses. Using this new idea, Padmanabhan \cite{Pad1} derived
the Friedmann equation of a flat Friedmann-Robertson-Walker (FRW) Universe.
Following \cite{Pad1}, further investigations have been carried out to
extract the Friedmann equations of a FRW Universe in various gravity
theories \cite{Cai1,Yang,FQ,Shey1}. In these investigations (\cite%
{Cai1,Yang,Shey1,FQ}), following \cite{Pad1}, the authors could
only derive the Friedmann equations of a flat FRW Universe and
they failed to obtain the dynamical equations describing the
evolution of the Universe with any spacial curvature in other
gravity theories. Very recently, an interesting modification of
Padmanabhan's proposal, which works in a nonflat Universe, was
suggested by Sheykhi \cite{Shey2}. Using this modified proposal
one is able to derive the corresponding dynamical equations
governing the evolution of the Universe with any spacial curvature
not only in Einstein gravity, but also in Gauss-Bonnet and more
general Lovelock gravity \cite{Shey2}. See also \cite{FF} for some
application and extension of \cite{Shey2}. In this paper, we will
address the question on the connection between the degrees of
freedom and the spacetime dynamics by investigating whether and
how the relation can be found in braneworld models.

Let us briefly review the proposal of \cite{Shey2}. According to Padmanabhan
in an infinitesimal interval $dt$ of cosmic time, the increase $dV$ of the
cosmic volume, in a flat Universe, is given by \cite{Pad1}
\begin{equation}
\frac{dV}{dt}=L_{p}^{2}(N_{\mathrm{sur}}-N_{\mathrm{bulk}}).  \label{dV1}
\end{equation}
where $L_{p}$ is the Planck length, $N_{\mathrm{sur}}$ is the number of
degrees of freedom on the boundary and $N_{\mathrm{bulk}}$ is the number of
degrees of freedom in the bulk. Through this paper we set $k_{B}=1=c=\hbar $
for simplicity. Inspired by (\ref{dV1}), an improved extension for $%
n\geq4 $-dimensional Universe with spacial curvature was found as \cite%
{Shey2}
\begin{equation}
\beta \frac{dV}{dt}=L_{p}^{n-2}H\tilde{r}_{A}\left(N_{\mathrm{sur}}-N_{%
\mathrm{bulk}}\right),  \label{dV21}
\end{equation}%
where $H=\dot{a}/a$ is the Hubble parameter, $a$ is the scale factor, $%
\beta ={(n-2)}/{2(n-3)}$ and $\tilde{r}_A$ is the apparent horizon
radius of FRW Universe given by
\begin{equation}  \label{radius}
\tilde{r}_A=\frac{1}{\sqrt{H^2+k/a^2}}.
\end{equation}
Motivated by the area law of the entropy, we assume the number of degrees of
freedom on the apparent horizon is
\begin{equation}  \label{Nsur1}
N_{\mathrm{sur}}=\beta \frac{A}{L_{p}^{n-2}},
\end{equation}
where $A=(n-1)\Omega _{n-1}\tilde{r}_{A}^{n-2}$ is the area of the apparent
horizon with $\Omega_{n-1}$ is the volume of a unit $(n-1)$-sphere. The
volume of the $(n-1)$-sphere with radius $\tilde{r}_{A}$ is $V =\Omega_{n-1}
\tilde{r}_{A}^{n-1}$. We assume the energy content inside the $n$%
-dimensional bulk is in the form of Komar energy \cite{Cai1}
\begin{equation}
E_{\mathrm{Komar}}=\frac{(n-3)\rho+(n-1)p}{n-3}V,  \label{Ek2}
\end{equation}
where $\rho$ and $p$ are the energy density and pressure of
the perfect fluid inside the Universe, respectively. Hence according to the equipartition
law of energy, the bulk degrees of freedom is obtained as
\begin{eqnarray}  \label{Nbulk1}
N_{\mathrm{bulk}}&=&\frac{2\left\vert E_{\mathrm{komar}}\right\vert }{T}
\notag \\
&=&-4 \pi \Omega_{n-1} \tilde{r}^{n}_A \frac{(n-3)\rho+(n-1)p}{n-3},
\end{eqnarray}
where $T=1/(2\pi \tilde{r}_{A})$ is the Hawking temperature associated with
the apparent horizon. Substituting Eqs. (\ref{Nsur1}) and (\ref{Nbulk1}) in
relation (\ref{dV2}), we arrive at
\begin{equation}  \label{Fr3}
H^{-1}\dot{r}_{A}\tilde{r}_{A}^{-3}-\tilde{r}_{A}^{-2}=\frac{8\pi
L_{p}^{n-2} }{(n-1)}\times\frac{(n-3)\rho +(n-1)p}{(n-2)}
\end{equation}%
Multiplying both hand sides of by factor $2\dot{a}a$, and using the
$n$-dimensional continuity equation:
\begin{equation}
\dot{\rho}+(n-1)H(\rho +p)=0,  \label{cont}
\end{equation}
we obtain \cite{Shey2}
\begin{equation}
\frac{d}{dt}\left[a^2 \left(H^2+\frac{k}{a^2}\right)\right]=\frac{16 \pi
L_{p}^{n-2}}{(n-1)(n-2)} \frac{d}{dt}(\rho a^2).  \label{Fr4}
\end{equation}
After integrating and setting the constant of integration equal to zero, we
find
\begin{equation}
H^2+\frac{k}{a^2}=\frac{16 \pi L_{p}^{n-2}}{(n-1)(n-2)}\rho.  \label{FrE}
\end{equation}
This is the Friedmann equation of $n$-dimensional FRW Universe with any
spacial curvature \cite{CaiKim}.

\section{Emergence of Friedmann equations in RS II braneworld}

\label{brane} In the remaining part of paper, we want to extend
the study to the branworld scenarios. Gravity on the brane does
not obey Einstein theory, thus the usual area formula for the
holographic boundary get modified on the brane
\cite{SheyW1,SheyW2}. Two well-known scenarios in braneworld are
Randall-Sundrum (RS) II \cite{RS,Bin} and Dvali, Gabadadze,
Porrati (DGP) \cite{DGP,DG} models. In the first scenario an
$(n-1)$-dimensional brane embedded in an $(n+1)$-dimensional AdS
bulk. In this case, the extra dimension has a finite size and the
localization of gravity on the brane occurs due to the negative
cosmological constant in the bulk. In the second scenario which is
called DGP model, an $(n-1)$-dimensional brane is embedded in a
spacetime with an infinite-size extra dimension, with the hope
that this picture could shed new light on the standing problem of
the cosmological constant as well as on supersymmetry breaking
\cite{DGP}. In the original DGP model the bulk was assumed to be a
Minkowskian spacetime with infinite size. In this case the
recovery of the usual gravitational laws on the brane is obtained
by adding an Einstein-Hilbert term to the action of the brane
computed with the brane intrinsic curvature. The so-called warped
DGP model corresponds to the case where both the intrinsic
curvature term on the brane and the negative cosmological constant
in the bulk are taken into account.

In order to apply the proposal (\ref{dV2}) to braneworld scenarios, we
modify it a little by replacing $L_{p}^{n-2}$ with $G_{n+1}$, namely
\begin{equation}
\beta \frac{dV}{dt}=G_{n+1}H\tilde{r}_{A}\left( N_{\mathrm{sur}}-N_{\mathrm{%
bulk}}\right) .  \label{dV2}
\end{equation}%
First of all, we consider the RS II scenario. The apparent horizon entropy
for an $(n-1)$-brane embedded in an $(n+1)$-dimensional bulk in RS II model
is given by \cite{SheyW1}
\begin{equation}
S=\frac{2\Omega _{n-1}{\tilde{r}_{A}}^{n-1}}{4G_{n+1}}\times
{}_{2}F_{1}\left( \frac{n-1}{2},\frac{1}{2},\frac{n+1}{2},-\frac{{\tilde{r}%
_{A}}^{2}}{\ell ^{2}}\right) ,  \label{entRSAdS1}
\end{equation}%
where ${}_{2}F_{1}(a,b,c,z)$ is a hypergeometric function, and $\ell $ is
the bulk AdS radius,
\begin{equation}
\ell ^{2}=-\frac{n(n-1)}{16\pi G_{n+1}\Lambda _{n+1}}\,,\quad \Omega _{n-1}=%
\frac{\pi ^{(n-1)/2}}{\Gamma ((n+1)/2)}.  \label{rela}
\end{equation}%
In the above relation, $\Lambda _{n+1}$ represents the $(n+1)$-dimensional
bulk cosmological constant. The entropy expression (\ref{entRSAdS1}) can be
written in the form \cite{SheyW1}
\begin{equation}
S=\frac{(n-1)\ell \Omega _{n-1}}{2G_{n+1}}{\displaystyle\int_{0}^{\tilde{r}%
_{A}}\frac{\tilde{r}_{A}^{n-2}}{\sqrt{\tilde{r}_{A}^{2}+\ell ^{2}}}d\tilde{r}%
_{A}},  \label{entRSAdS2}
\end{equation}%
and hence we define the effective area as
\begin{equation}
\widetilde{A}=4G_{n+1}S=2(n-1)\ell \Omega _{n-1}\int_{0}^{\tilde{r}_{A}}%
\frac{\widetilde{r}_{A}^{n-2}}{\sqrt{\widetilde{r}_{A}^{2}+\ell ^{2}}}d%
\widetilde{r}_{A}^{{}}
\end{equation}%
Now we calculate the increasing in the effective volume as
\begin{eqnarray}
\frac{d\widetilde{V}}{dt} &=&\frac{\tilde{r}_{A}}{(n-2)}\frac{d\tilde{A}}{dt}
\notag \\
&=&2\ell \Omega _{n-1}\frac{(n-1)}{(n-2)}\frac{\tilde{r}_{A}^{n-1}}{\sqrt{%
\tilde{r}_{A}^{2}+\ell ^{2}}}\dot{\tilde{r}}_{A} \\
&=&-2\Omega _{n-1}\frac{(n-1)}{(n-2)}\tilde{r}_{A}^{n+1}\frac{d}{dt}\left(
\sqrt{\tilde{r}_{A}^{-2}+\frac{1}{\ell ^{2}}}\right)  \label{dVt1}
\end{eqnarray}%
Motivated by (\ref{dVt1}), we assume the number of degrees of freedom on the
boundary is given by
\begin{eqnarray}
N_{\mathrm{sur}} &=&\frac{4\beta (n-1)\Omega _{n-1}}{(n-2)G_{n+1}}\tilde{r}%
_{A}^{n}\sqrt{\tilde{r}_{A}^{-2}+\frac{1}{\ell ^{2}}}\nonumber \\
&=&\frac{2(n-1)\Omega _{n-1}}{(n-3)G_{n+1}}\tilde{r}_{A}^{n}\sqrt{\tilde{r}%
_{A}^{-2}+\frac{1}{\ell ^{2}}}. \label{NsurRS}
\end{eqnarray}%
Inserting Eqs. (\ref{Nbulk1}), (\ref{dVt1}) and (\ref{NsurRS}) in relation (%
\ref{dV2}), after multiplying both hand side by factor $\dot{a}a$,
we get
\begin{eqnarray}
&&-\frac{\widetilde{r}_{A}^{-3}\dot{\widetilde{r}}_{A}}{\sqrt{\widetilde{r}%
_{A}^{-2}+\frac{1}{\ell ^{2}}}}a^{2}+2\dot{a}a\sqrt{\widetilde{r}_{A}^{-2}+%
\frac{1}{\ell ^{2}}}  \notag \\
&=&-4\pi G_{n+1}\dot{a}a\left( \frac{(n-3)\rho +(n-1)p}{(n-1)}\right) .
\end{eqnarray}%
Using the continuity equation (\ref{cont}), after some simplification, we
arrive at
\begin{equation}
\frac{d}{dt}\left( a^{2}\sqrt{\tilde{r}_{A}^{-2}+\frac{1}{\ell ^{2}}}\right)
=\frac{4\pi G_{n+1}}{(n-1)}\frac{d}{dt}\left( \rho a^{2}\right) .
\end{equation}%
Integrating and dividing by $a^{2}$, we find
\begin{equation}
\sqrt{\tilde{r}_{A}^{-2}+\frac{1}{\ell ^{2}}}=\frac{4\pi G_{n+1}}{(n-1)}\rho
,  \label{FRS}
\end{equation}%
where we assumed the integration constant to be zero. Substituting the
apparent horizon radius from relation (\ref{radius}), we get
\begin{equation}
\sqrt{H^{2}+\frac{k}{a^{2}}+\frac{1}{\ell ^{2}}}=\frac{4\pi G_{n+1}}{(n-1)}%
\rho .  \label{FrRS}
\end{equation}%
In this way we derive the Friedmann equation of higher dimensional
FRW Universe in RS II braneworld by calculating the difference
between the number of degrees of freedom on the boundary and in
the bulk. This coincides with the result obtained in \cite{SheyW1}
from the field equations.

\section{Friedmann equations in Warped DGP braneworld}
Next we consider an $(n-1)$-dimensional warped DGP brane embedded in an $%
(n+1)$-dimensional AdS bulk. The entropy associated with the apparent
horizon is given by \cite{SheyW1}
\begin{eqnarray}
S&=&\frac{(n-1)\Omega _{n-1}{\tilde{r}_{A}}^{n-2}}{4G_{n}}+\frac{2\Omega
_{n-1}{\tilde{r}_{A}}^{n-1}}{4G_{n+1}}  \notag \\
&&\times {}_{2}F_{1}\left( \frac{n-1}{2},\frac{1}{2},\frac{n+1}{2},-\frac{{%
\tilde{r}_{A}}^{2}}{\ell ^{2}}\right).  \label{entDGP}
\end{eqnarray}%
It is important to note that in DGP braneworld, the entropy
expression of the apparent horizon consists two terms. The first term
which satisfies the area formula on the brane is the
contribution from the Einstein-Hilbert term on the brane. The
second term is the same as the entropy of RS II braneword
and therefore obeys the $(n+1)$-dimensional area law in the bulk
\cite{SheyW1}.

One can write the entropy associated with the apparent horizon on the brane
as \cite{SheyW1}
\begin{equation}
S=(n-1)\Omega _{n-1}\int_{0}^{\tilde{r}_{A}}\left( \frac{(n-2)\tilde{r}%
_{A}^{n-3}}{4G_{n}}+\frac{\ell }{2G_{n+1}}\frac{\tilde{r}_{A}^{n-2}}{\sqrt{%
\tilde{r}_{A}^{2}+\ell ^{2}}}\right) d\tilde{r}_{A}^{{}}
\end{equation}%
We define the effective surface as
\begin{eqnarray}
\widetilde{A} &=&4G_{n+1}S=4G_{n+1}(n-1)\Omega _{n-1}  \notag \\
&&\times \int_{0}^{\tilde{r}_{A}}\left( \frac{(n-2)\tilde{r}_{A}^{n-3}}{%
4G_{n}}+\frac{\ell }{2G_{n+1}}\frac{\tilde{r}_{A}^{n-2}}{\sqrt{\tilde{r}%
_{A}^{2}+\ell ^{2}}}\right) d\tilde{r}_{A}.  \notag \\
&&
\end{eqnarray}%
We also obtain the rate of increase in the effective volume as
\begin{eqnarray}
\frac{d\widetilde{V}}{dt} &=&\frac{\tilde{r}_{A}}{(n-2)}\frac{d\tilde{A}}{dt}%
=\Omega _{n-1}\frac{(n-1)}{(n-2)}\dot{\tilde{r}}_{A}\tilde{r}_{A}^{n-2}
\notag \\
&&\times \left( \frac{(n-2)G_{n+1}}{G_{n}}+\frac{2}{\sqrt{\tilde{r}%
_{A}^{-2}+\ell ^{-2}}}\right)  \notag \\
&=&-2\Omega _{n-1}\frac{(n-1)}{(n-2)}\tilde{r}_{A}^{n+1}  \notag \\
&&\times \frac{d}{dt}\left( \frac{(n-2)G_{n+1}}{4G_{n}}\tilde{r}_{A}^{-2}+%
\sqrt{\tilde{r}_{A}^{-2}+\frac{1}{\ell ^{2}}}\right)  \label{dVt2}
\end{eqnarray}%
Inspired by (\ref{dVt2}), we suppose the number of degrees of freedom on the
apparent horizon in warped DGP model is given by
\begin{eqnarray}
N_{\mathrm{sur}}=\frac{2\Omega _{n-1}}{G_{n+1}}\frac{(n-1)}{(n-3)}\tilde{r}%
_{A}^{n} &&\left( \frac{G_{n+1}(n-2)\tilde{r}_{A}^{-2}}{4G_{n}}\right.
\notag \\
&&\left. +\sqrt{\tilde{r}_{A}^{-2}+\frac{1}{\ell ^{2}}}\right) .
\label{NsurDGP}
\end{eqnarray}%
Combining Eqs. (\ref{Nbulk1}), (\ref{dVt2}) and (\ref{NsurDGP}) with
relation (\ref{dV2}), it is a matter of calculation to find
\begin{eqnarray}
\frac{d}{dt}\left( a^{2}\sqrt{\tilde{r}_{A}^{-2}+\frac{1}{\ell ^{2}}}\right)
&=&-\frac{G_{n+1}}{4G_{n}}(n-2)\frac{d}{dt}\left( \tilde{r}%
_{A}^{-2}a^{2}\right)  \notag \\
&&+\frac{4\pi G_{n+1}}{(n-1)}\frac{d}{dt}\left( \rho a^{2}\right) .
\end{eqnarray}%
Integrating and dividing by $a^{2}$ we obtain
\begin{equation}
\sqrt{\tilde{r}_{A}^{-2}+\frac{1}{\ell ^{2}}}=-\frac{G_{n+1}}{4G_{n}}(n-2)%
\tilde{r}_{A}^{-2}+\frac{4\pi G_{n+1}}{(n-1)}\rho .
\end{equation}%
Substituting the apparent horizon radius from relation (\ref{radius}), we
have
\begin{eqnarray}
&&\sqrt{H^{2}+\frac{k}{a^{2}}+\frac{1}{\ell ^{2}}}+\frac{G_{n+1}}{4G_{n}}%
(n-2)\left( H^{2}+\frac{k}{a^{2}}\right)  \notag  \label{FrDGP} \\
&=&\frac{4\pi G_{n+1}}{(n-1)}\rho .
\end{eqnarray}%
This equation is indeed the Friedmann equation of FRW Universe in warped DGP
braneworld derived in \cite{SheyW1} from the field equations. If we define,
as usual, the crossover length scale between the small and large distances
in DGP braneworld as \cite{Def}
\begin{equation}
r_{c}=\frac{G_{n+1}}{2G_{n}},
\end{equation}%
then one can easily check that for $r_{c}\rightarrow \infty $, the standard
Friedmann equation in $n$-dimensional FRW Universe presented in (\ref{FrE})
is recovered. On the other hand, when $r_{c}\rightarrow 0$, Eq. (\ref{FrDGP}%
) reduces to the Friedmann equation in RS II braneworld obtained in the
previous section.

\section{Emergence of spacetime dynamics in Gauss-Bonnet braneworld}

Finally, we apply the method developed in the previous sections to
investigate the emergence properties of the spacetime dynamics in
general braneworld with curvature correction terms including a 4D
scalar curvature from induced gravity on the brane, and a 5D
Gauss-Bonnet curvature term in the bulk. With these correction
terms, especially including a Gauss-Bonnet correction to the 5D
action, we have the most general action with second-order field
equations in 5D \cite{lovelock}, which provides the most general
models for the braneworld scenarios. The entropy of apparent
horizon in general Gauss-Bonnet braneworld embedded in a 5D bulk,
can be written as \cite{SheyW2}
\begin{eqnarray}
S &=&\frac{3\Omega _{3}{\tilde{r}_{A}}^{2}}{4G_{4}}+\frac{2\Omega _{3}{%
\tilde{r}_{A}}^{3}}{4G_{5}}\times {}_{2}F_{1}\left( \frac{3}{2},\frac{1}{2},%
\frac{5}{2},\Phi _{0}{\tilde{r}_{A}}^{2}\right)   \notag  \label{ent2} \\
&&+\frac{6{\alpha}\Omega _{3}{\tilde{r}_{A}}^{3}}{G_{5}}\left(
\Phi
_{0}\times {}_{2}F_{1}\left( \frac{3}{2},\frac{1}{2},\frac{5}{2},\Phi _{0}{%
\tilde{r}_{A}}^{2}\right) \right.   \notag \\
&&\left. +\frac{\sqrt{1-\Phi _{0}\tilde{r}_{A}^{2}}}{{\tilde{r}_{A}}^{2}}%
\right) ,
\end{eqnarray}%
where $\Phi _{0}=\frac{1}{4{\alpha}}\left( -1+\sqrt{1-\frac{8{%
\alpha}}{\ell ^{2}}}\right) $=constant \cite{SheyW2}, and
${\alpha}$ is the Gauss-Bonnet coefficient with dimension (length)$^{2}$. When ${%
\alpha}\rightarrow 0$ we have $\Phi _{0}=-\ell ^{-2}$ and the above
expression reduces to the entropy of warped DGP braneworld presented in (\ref%
{entDGP}) for $n=4$. Expression (\ref{ent2}) can be written as
\cite{SheyW2}
\begin{eqnarray}
S &=&\frac{3\Omega _{3}}{2G_{4}}\int_{0}^{\tilde{r}_{A}}\tilde{r}_{A}d\tilde{%
r}_{A}+\frac{3\Omega _{3}}{2G_{5}}\int_{0}^{\tilde{r}_{A}}\frac{\tilde{r}%
_{A}^{2}}{\sqrt{1-\Phi _{0}\tilde{r}_{A}^{2}}}d\tilde{r}_{A}  \notag \\
&&+\frac{6{\alpha}\Omega _{3}}{G_{5}}\int_{0}^{\tilde{r}_{A}}\frac{%
2-\Phi _{0}\tilde{r}_{A}^{2}}{\sqrt{1-\Phi _{0}\tilde{r}_{A}^{2}}}d\tilde{r}%
_{A},
\end{eqnarray}%
We define the effective area of the apparent horizon corresponding to the
above entropy as
\begin{eqnarray}
\tilde{A}=4G_{5}S &=&\frac{6G_{5}\Omega _{3}}{G_{4}}\int_{0}^{\tilde{r}_{A}}%
\tilde{r}_{A}d\tilde{r}_{A}+6\Omega _{3}\int_{0}^{\tilde{r}_{A}}\frac{\tilde{%
r}_{A}d\tilde{r}_{A}}{\sqrt{\tilde{r}_{A}^{-2}-\Phi _{0}}}  \notag \\
&&+24{\alpha}\Omega _{3}\int_{0}^{\tilde{r}_{A}}\frac{2\tilde{r}%
_{A}^{-1}-\Phi _{0}\tilde{r}_{A}}{\sqrt{\tilde{r}_{A}^{-2}-\Phi _{0}}}d%
\tilde{r}_{A},
\end{eqnarray}%
and therefore the increase of the effective volume is obtained as
\begin{eqnarray}
\frac{d\widetilde{V}}{dt}=\frac{\tilde{r}_{A}}{2}\frac{d\tilde{A}}{dt} &=&%
\frac{3G_{5}\Omega _{3}}{G_{4}}\tilde{r}_{A}^{2}\dot{\tilde{r}}_{A}+3\Omega
_{3}\frac{\tilde{r}_{A}^{2}\dot{\tilde{%
r}}_{A} }{\sqrt{\tilde{r}_{A}^{-2}-\Phi _{0}}} \notag \\
&&+12{\alpha}\Omega _{3}\frac{2-\Phi _{0}\tilde{r}_{A}^{2}}{\sqrt{%
\tilde{r}_{A}^{-2}-\Phi _{0}}}\dot{\tilde{r}}_{A}.  \label{dVt3}
\end{eqnarray}%
Motivated by (\ref{dVt3}), we write the number of degrees of freedom on the
boundary in general Gauss-Bonnet braneworld as
\begin{eqnarray}
N_{\mathrm{sur}} &=&\frac{3\Omega _{3}}{G_{4}}\tilde{r}_{A}^{2}+\frac{%
6\Omega _{3}}{G_{5}}\tilde{r}_{A}^{4}\sqrt{\tilde{r}_{A}^{-2}-\Phi _{0}}
\notag \\
&&+\frac{16{\alpha}\Omega _{3}}{G_{5}}\tilde{r}_{A}^{4}\left( \tilde{r}%
_{A}^{-2}+\frac{\Phi _{0}}{2}\right) \sqrt{\tilde{r}_{A}^{-2}-\Phi _{0}}.
\label{NsurGB}
\end{eqnarray}%
Substituting Eqs. (\ref{Nbulk1}), (\ref{dVt3}) and (\ref{NsurGB}) into (\ref
{dV2}) and setting $n=4$, after some mathematic simplification, one obtains
\begin{eqnarray}
&&\frac{3G_{5}}{G_{4}}\frac{d}{dt}\left( a^{2}\tilde{r}_{A}^{-2}\right) +6%
\frac{d}{dt}\left( a^{2}\sqrt{\tilde{r}_{A}^{-2}-\Phi _{0}}\right)   \notag
\\
&&+\frac{d}{dt}\Bigg{\{}16{\alpha}a^{2}\left( \tilde{r}_{A}^{-2}+\frac{%
\Phi _{0}}{2}\right) \sqrt{\tilde{r}_{A}^{-2}-\Phi _{0}}\Bigg{\}}  \notag \\
&=&8\pi G_{5}\frac{d}{dt}\left( \rho a^{2}\right) .
\end{eqnarray}%
Integrating, dividing by $a^{2}$ and then using the definition (\ref{radius}%
), we find
\begin{eqnarray}
&&\left[ 1+\frac{8}{3}{\alpha}\left( H^{2}+\frac{k}{a^{2}}-\frac{1}{%
2\ell ^{2}}\right) \right] \sqrt{H^{2}+\frac{k}{a^{2}}+\frac{1}{\ell ^{2}}}
\notag \\
&=&\frac{4\pi G_{5}}{3}\rho -\frac{G_{5}}{2G_{4}}\left( H^{2}+\frac{k}{a^{2}}%
\right) .
\end{eqnarray}%
This is the Friedmann equation governing the evolution of the
Universe in general Gauss-Bonnet braneworld with curvature
correction terms on the brane and in the bulk. This result is
exactly the same as one obtains from the field equation of
Gauss-Bonnet braneworld \cite{kofin}. Here we arrived at
the same result by using the novel proposal of \cite{Shey2}. When $\alpha =0$%
, the above result reduces to the Friedmann equation of warped DGP model
obtained in Eq. (\ref{FrDGP}) for $n=4$.

\section{Summery and discussion\label{Con}}

Recently, Padmanabhan \cite{Pad1} argued that the spacetime
dynamics can be considered as an emergent phenomena and the cosmic
space is emergent as the cosmic time progresses. An improved
version of Padmanabhan proposal which is applicable to a nonflat
Universe was found by one of the present author \cite{Shey2}. In
this paper, we extended the study to other gravity theory such as
braneworld scenarios. Gravity on the brane does not obey the
Einstein theory and therefore the usual area formula for the
entropy does not hold on the brane. We have discussed several
cases including whether there is or not a Gauss-Bonnet curvature
correction term in the bulk and whether there is or not an
intrinsic curvature term on the brane. We found that one can
always derive the Friedmann equations of FRW Universe with any
spacial curvature, by calculating the difference between the
horizon degrees of freedom and the bulk degrees of freedom
regardless of the existence of the intrinsic curvature term
on the brane and the Gauss-Bonnet correction term in the bulk.

The result obtained here in RS II, warped DGP and the general Gauss-Bonnet
braneworld scenarios further supports the novel idea of Padmanabhan (\ref%
{dV1}) and its extension as (\ref{dV2}), and show that this
approach is powerful enough to extract the dynamical equations
describing the evolution of the Universe in other gravity theories
with any spacial curvature.

\acknowledgments{We thank from the Research Council of Shiraz
University. This work has been supported financially by Research
Institute for Astronomy \& Astrophysics of Maragha (RIAAM), Iran}.

\end{document}